\documentclass[%
superscriptaddress,
showpacs,
amsmath,amssymb,
aps,
prc,
showkeys,
twocolumn,
]{revtex4-1}

\usepackage{amsmath,mathrsfs,amssymb}
\usepackage{eurosym}
\usepackage{graphicx,subfigure}
\usepackage{color}
\usepackage{indentfirst}
\usepackage{extarrows}
\usepackage{hyperref}
\hypersetup{colorlinks=true, citecolor=blue, urlcolor=blue, linkcolor=blue}

\begin{document}

\title{Astrophysical factors of ${}^{12}{\rm C}+{}^{12}{\rm C}$ fusion from Trojan horse method}
\author{A. M. Mukhamedzhanov}
\email{akram@comp.tamu.edu}
\affiliation{Cyclotron Institute, Texas A$\&$M University, College Station, TX 77843, USA}
\author{D. Y. Pang}
\email{dypang@buaa.edu.cn}
\affiliation{School of Physics and Nuclear Energy Engineering, Beihang University, Beijing, 100191,People's Republic of China}
\affiliation{Beijing Key Laboratory of Advanced Nuclear Materials and Physics, Beihang University, Beijing 100191, China}

\begin{abstract}
Carbon-carbon burning plays an important role in many stellar environments. Recently, using the indirect Trojan horse method  A. Tumino {\em et al.} reported [Nature {\bf 557} 687 (2018)] a strong rise of the astrophysical factor for the carbon-carbon fusion  at low resonance energies. In this paper, we demonstrate that this rise is the artifact of using an invalid 
plane-wave approximation.  It is shown that the calculated  renormalization factor decreases the astrophysical factor from  [A. Tumino {\em et al.}, Nature {\bf 557} 687 (2018)]
at the resonance energies of $E=0.8-0.9$ MeV by as much as $\approx 10^{3}$ times.
.
\end{abstract}

\maketitle

\date{Today}


Recently the indirect Trojan horse method (THM) was applied to measure the astrophysical $S^\ast$-factor of the ${}^{12}{\rm C}-{}^{12}{\rm C}$ fusion \cite{Nature}. In the THM a surrogate reaction $a(sx)+ A \to s+ F(xA) \to s+ b+ B$ is used to determine the astrophysical $S^\ast(E)$-factor of the binary resonant sub-reaction $\,x+A \to F \to b+B$. In the case under consideration $a={}^{14}{\rm N}$, $A={}^{12}{\rm C},$ $\;x={}^{12}{\rm C}$, $s=d$, and $F={}^{24}{\rm Mg}^\ast$. Four different channels in the final state were populated in the THM experiment: $p_{0} + {}^{23}{\rm Na}$, $p_{1} + {}^{23}{\rm Na}$  (0.44 \, MeV), $\alpha_{0} + {}^{20}{\rm Ne}$, and $\alpha_{1} + {}^{20}{\rm Ne}$ (1.63\, MeV) \cite{Nature}. 

 A simple plane-wave approximation (PWA) was used in  \cite{Nature} to analyze the THM data. This PWA follows from a more general expression, which contains the distorted waves in the initial and final states [see Eq. (117) of Ref. \cite{muk2011}].  A generalized $R$-matrix approach  was developed in Ref. \cite{muk2011} using the surface integral formalism. The approach is suitable for the analysis of $2\, {\it particles} \to 3\, {\it particles}$ reactions proceeding through an intermediate resonance in the binary subsystem.  To simplify the theory presented in \cite{muk2011}, one of us (A.M.M., who is also the author of \cite{muk2011}), developed the aforementioned PWA \cite{reviewpaper}. This approximation can be applied for the analysis of the THM reactions in which the spectator is a neutron or for reactions at energies above the Coulomb barrier in the initial and final states, and when the interacting nuclei have small charges.   It is assumed in the PWA that the angular distribution of the spectator is forward peaked in the center-of-mass system (quasi-free kinematics) and that the bound-state wave function of the spectator can be factorized out [see Eq. (117) of Ref. \cite{muk2011} and Eq. (2) of Ref. \cite{Nature}]. It was found in \cite{Nature} that the astrophysical factors extracted from the THM experiment demonstrate a steep rise when the resonance energy decreases. This rise would have profound implications on different astrophysical scenarios because the carbon-carbon fusion rate calculated from the astrophysical $S^\ast$-factors deduced in Ref.  \cite{Nature} significantly exceeds all the previous estimations of the reaction rate obtained by extrapolation of direct data to the low-energy region. For example, the reaction rate calculated in \cite{Nature} at tempetature  $T \sim 2 \times 10^{8}$ K exceeds the adopted value \cite{Fowler,Iliadis} by a factor of $500$.

Our goal is demonstrate that  the PWA cannot be used for the analysis of the THM $\,{}^{12}{\rm C}({}^{14}{\rm N},\,d,b)B\,$ reaction proceeding through the resonant states ${}^{24}{\rm Mg}^{*}$ in the intermediate binary subsystem, where particles $b$ and $B$ are $p$ and ${}^{23}{\rm Na}$ or $\alpha$ and ${}^{20}{\rm Ne}$ \cite{Nature}.
 
It is important to remind that in the THM only an energy dependence of the astrophysical factor is measured . Its absolute value is determined by normalizing  the THM data to the accurate direct data available at higher energies. In \cite{Nature} the normalization of the THM data to the direct data was done in the energy interval $\,E=2.5-2.63\,$ MeV, where $\,E_{{}^{12}{\rm C}\,{}^{12}{\rm C}}   \equiv E\,$  is the ${}^{12}{\rm C}- {}^{12}{\rm C}$ relative kinetic energy. To check whether the PWA is justified, we consider the kinematics of the THM in this energy interval. 

The THM experiment  was performed at the relative ${}^{14}{\rm N}-{}^{12}{\rm C}$ energy of $E_{aA}=13.845$ MeV \cite{Nature}. The energy conservation in the THM reaction requires that $E_{aA}+Q= E_{f}$, where  $\,Q= m_{a} + m_{A} - m_{s} - m_{b} - m_{B},$ $\,E_{f}=E_{sF}+E_{bB}$ is the total kinetic energy of the final three-body system $s+b+B$, $E_{ij}$ is the relative kinetic energy of the particles $i$ and $j$, $\;m_{i}$ is the mass of particle $i$. From this equation we get that for the final  $\,d+p + {}^{23}{\rm Na}$ channel the total kinetic energy is $\,E_{f}=5.8$ MeV. 

Let us consider the ${}^{24}{\rm Mg}^{*}$ resonance in the  ${}^{12}{\rm C}-{}^{12}{\rm C}$ channel at $E_{(R){}^{12}{\rm C}{}^{12}{\rm C}}^{(0)}=2.664$ MeV \cite{Nature}.
We use the notation: $E_{(R)(xA)}= E_{(R)xA}^{(0)} - i\,\Gamma/2$ stands for the complex resonance energy in the channel $x+A,\,$  $\,E_{(R)xA}^{(0)}$ is its real part and $\Gamma$ is the total width of the resonance. 
 . We assume that the THM reaction proceeds  as the two-step  process described by the diagram in Fig. (2) of Ref. \cite{reviewpaper}.
Taking into account that for the binary reaction $\,{}^{12}{\rm C} + {}^{12}{\rm C} \to p+{}^{23}{\rm Na}$, $Q_{2}=2.24$ MeV, where $Q_{2} =m_{x} + m_{A} - m_{b}- m_{B}$, we get that the resonance energy in the $\,p+{}^{23}{\rm Na}$ channel corresponding to $E_{(R){}^{12}{\rm C}{}^{12}{\rm C}}^{(0)}=2.664$ MeV is $E_{(R)p{}^{23}{\rm Na}}^{(0)}=4.9$ MeV. Hence, the relative kinetic energy of the deuteron and the c.m. of the $\,p + {}^{23}{\rm Na}$ system corresponding to this resonance is $E_{d{}^{24}{\rm Mg}}=0.9$ MeV.  This energy is well bellow the Coulomb barrier in the $d$-$^{24}{\rm Mg}$ system, which is about 3 MeV. Even on the lower end of the normalization interval corresponding to $E=2.5$ MeV, the relative energy is 
$E_{d{}^{24}{\rm Mg}}=1.06$ MeV.  

At the resonance energy of $\,E_{(R){}^{12}{\rm C}{}^{12}{\rm C}}^{(0)}=1.5$ MeV in the $\,{}^{12}{\rm C}-{}^{12}{\rm C}$ channel, which corresponds to the resonance  energy $\,E_{(R)p{}^{23}{\rm Na}}^{(0)}= 3.74$ MeV in the exit channel, the relative kinetic energy $\,E_{d{}^{24}{\rm Mg}}=2.06$ MeV, which is still below the Coulomb barrier. Note that the resonance energies, which can be observed in the THM experiment, are $E < 3.56$ MeV, because at $E >3.56$ MeV, the resonance energy in the $p+{}^{23}{\rm Na}$ channel is $3.56+Q_{2}>5.8$ MeV, that is, the $d-{}^{24}{\rm Mg}$ relative kinetic energy is $E_{d-{}^{24}{\rm Mg}}<0$. That is why the extrapolation of the $S^\ast(E)$-factor beyond this energy, as it is done in \cite{arxivAurora}, is difficult to justify.

Even for the resonance energy of $\,E_{(R){}^{12}{\rm C}{}^{12}{\rm C}}^{(0)}=0.805$ MeV, which corresponds to $E_{(R)p{}^{23}{\rm Na}}^{(0)}= 3.05$ MeV, the $d-{}^{24}{\rm Mg}$ relative kinetic energy is $E_{d{}^{24}{\rm Mg}}=2.75$ MeV, which is close to but still below the Coulomb barrier. Nevertheless, for the three-body system $d+ p+ {}^{23}{\rm Na}$ containing the nucleus $^{23}$Na with the charge $Z_{{}^{23}{\rm Na}}=11$, the Coulomb interaction plays a significant role. Presumably, the momentum distribution of the deuterons shown in \cite{Nature} was measured for the resonance energy in the ${}^{12}{\rm C}-{}^{12}{\rm C}$ channel close to $0.8$ MeV. Similar considerations can be done for other channels.

The presence of  strong Coulomb interactions for such deep sub-Coulomb processes in the final state of the transfer reaction significantly increases the differential cross section 
(DCS) in the backward hemisphere, shifting the peak of the angular distribution of the deuterons to the backward angles.  It completely contradicts to the PWA DCS in the c.m. system, which has a pronounced peak at forward angles. Even at the lowest observed resonances at $0.8-0.9$ MeV in the THM experiment \cite{Nature} the angular distribution of the deuterons noticeably deviates from the PWA one if the Coulomb (or Coulomb plus nuclear) rescattering effects in the initial and final states of the ${}^{12}{\rm C}$ transfer reaction are included.

But what is  more important is the fact that the presence of the strong Coulomb interactions significantly changes the absolute value and the energy
dependence of the DCS of the ${}^{12}{\rm C}$ transfer reaction. The absolute value of the DCS in the THM normalization interval becomes smaller than the corresponding PWA one by more than three orders of magnitude. It increases rapidly when the resonance energy decreases. That is  the  reason
 for the drop of the THM astrophysical factors found in this work compared to those extracted in the PWA \cite{Nature}. 
\begin{figure}[htbp]
\includegraphics[width=0.5\textwidth]{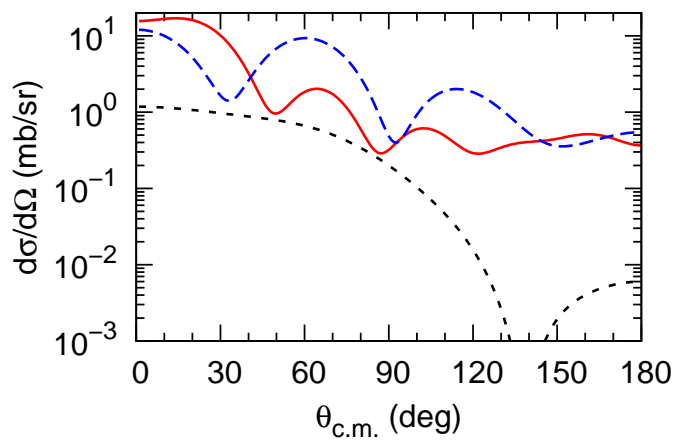}
\caption{The PWA DCSs calculated for the relative kinetic energy $E_{{}^{12}{\rm C}-{}^{14}{\rm N}}=13.85$ MeV  for the ${}^{14}{\rm N} + {}^{12}{\rm C} \to  d+ {}^{24}{\rm Mg}^{*}$ reaction populating three different resonant states. The black dotted, blue dashed , and red solid curves correspond to the resonant energies $E=2.7$, $1.5$ and $0.8$ MeV, respectively.}
\label{fig_PWA1}
\end{figure}

\begin{figure}[htbp]
\includegraphics[width=0.5\textwidth]{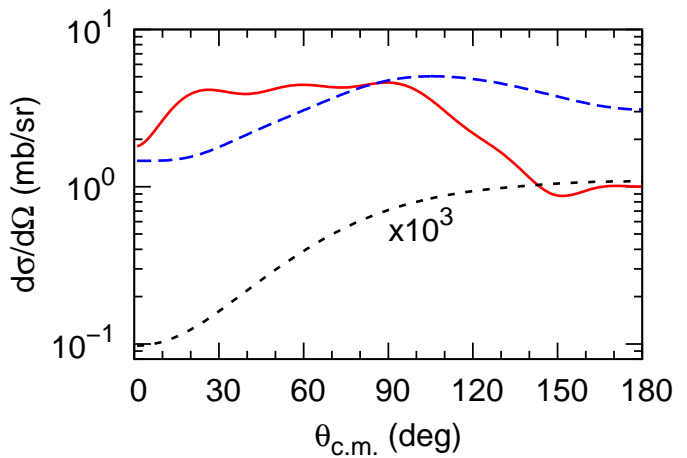}
\caption{The same as in Fig. \ref{fig_PWA1}, but the DCSs are calculated using the DWBA. Only the Coulomb distorted waves in the initial and final channels are included. Note that the DCS for $E=2.7$ MeV is multiplied by a factor of $10^{3}$.}
\label{fig_DWBAC1}
\end{figure}

\begin{figure}[htbp]
\includegraphics[width=0.5\textwidth]{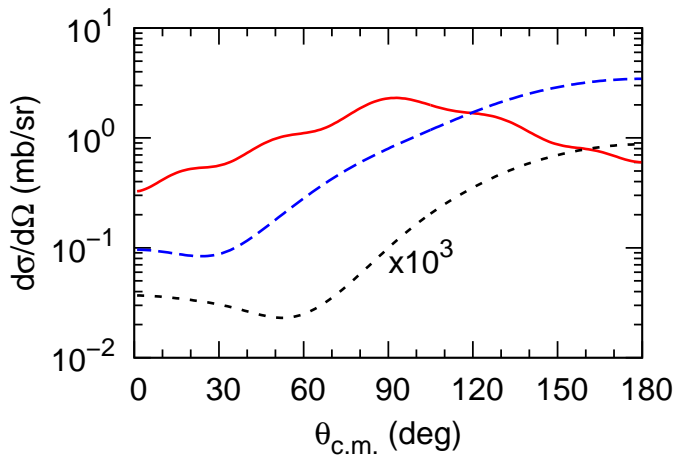}
\caption{The DWBA DCSs calculated using the Coulomb plus nuclear distorted waves. 
The notations are the same as in Fig. \ref{fig_DWBAC1}.} 
\label{fig_DWBACN1}
\end{figure}

\begin{figure}[htbp]
\includegraphics[width=0.5\textwidth]{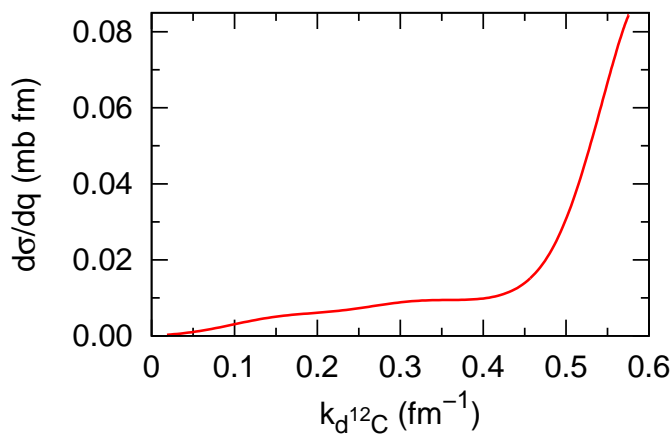}
\caption{DWBA DCS as function of the  $d-{}^{12}{\rm C}$ relative momentum at $E=2.7$ MeV. Only the Coulomb distorted waves are taken into account.}
\label{fig_momentdist}
\end{figure}

Figures \ref{fig_PWA1}-\ref{fig_momentdist} show the  DCSs in the c.m. frame for the ${}^{14}{\rm N} + {}^{12}{\rm C} \to  d+ {}^{24}{\rm Mg}^{*}$ reaction populating the resonant states $E=2.7$, 1.5, and 0.8 MeV. In Fig. \ref{fig_DWBAC1}  the DCSs were calculated using the distorted-wave Born approximation (DWBA) with pure Coulomb distorted waves in the initial and final states of the ${}^{12}{\rm C}$ transfer reaction. In Fig. \ref{fig_DWBACN1} the DWBA DCSs were calculated using the Coulomb plus nuclear distorted waves in the initial and final states.

 These figures  corroborate our statement that the PWA is invalid for the analysis of the THM reaction under consideration and allow us to draw the following compelling conclusions:
\begin{itemize}
\item
The angular distributions and absolute values of the PWA and  DWBA  DCSs differ significantly.   In particular, in the interval of the resonance  energies  $E=1.5-2.7$ MeV the DWBA DCSs calculated using the Coulomb distorted waves or the Coulomb plus nuclear distorted wave have minimum at forward angles. These DCSs increase when the scattering angle of the deuterons increases. It also demonstrates that the statement in \cite{arxivAurora}  that the deuterons cannot be emitted to backward angles
is not correct.
\item
The ratio of the DCSs from the PWA and the DWBA at $E=2.7$ MeV and $0.8$ MeV are completely different.  The DWBA DCSs at any angle at $E=2.7$ MeV are significantly smaller than those at $E=0.8$ MeV. This takes place if only the Coulomb or the Coulomb plus nuclear distorted waves are taken into account. This is an additional corroboration of the fact that at the resonance energies of the THM normalization interval $E=2.5-2.63$ MeV \cite{Nature} the THM reactions are deep sub-Coulomb. Hence, their DWBA DCSs are  extremely small. The absolute value of the DWBA DCS quickly increases when $E$ decreases because the energy of the outgoing deuteron increases approaching the Coulomb barrier. 

\item
In Eq. (\ref{SfactorarbE1}) (see below) for the $S$-factor the DWBA DCS appears in the denominator. A very small DCS at higher $E$ should significantly increase the THM astrophysical factor. As the energy $E$ decreases the DWBA DCS increases and the $S(E)$-factor quickly drops. For comparison, we normalized our renormalization factor $R(E)$, see below Eq. (\ref{RF1}), to unity at $E=2.664$ MeV, which is on the upper border of the THM normalization interval used in \cite{Nature}. The significant rise of the  DWBA DCS toward small $E$ is the factor that most contributes to the  drop of the THM $S(E)$-factor.

\item
The  $d-{}^{12}{\rm C}$ momentum distribution  at $E=2.7$ MeV is  opposite to the momentum distribution  given by the Fourier transform of the $(d-{}^{12}{\rm C})$ bound-state wave function in ${}^{14}{\rm N}$, see the extended data given in Fig. 1 of Ref. \cite{Nature}. This serves as an additional confirmation that the PWA-based Eq. (2) of Ref. \cite{Nature} for the reaction under consideration, which leads to the factorization of the $(d-{}^{12}{\rm C})$ bound-state wave function, cannot be used especially at higher resonance energies.
The PWA introduces the largest errors at higher resonance energies.  At these energies, the energies of the deuterons are lower, and the difference between the PWA and DWBA  is the largest.
No information in \cite{Nature} is available about the $d-{}^{12}{\rm C}$ relative momentum  distribution in the THM  normalization interval.
\end{itemize}

\begin{figure}[tbp] 
\includegraphics[width=4.0in,height=4.5in]{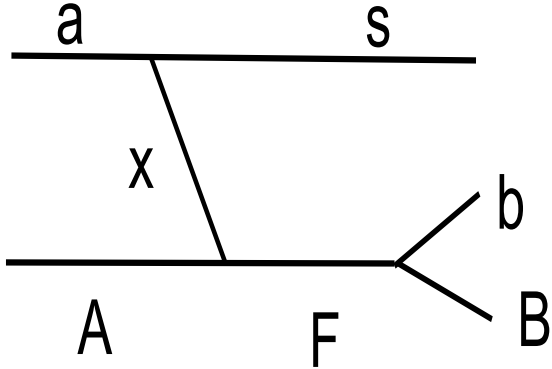}
  \caption{Simple diagram describing the THM mechanism in the plane-wave approximation}
\label{fig_THMPWAdiargam1}
\end{figure}

Having demonstrated that the Coulomb effects must be included, we briefly describe a correct procedure, which shows how the $S$-factors deduced in \cite{Nature} should be renormalized due to the Coulomb effects. 
The diagram shown in Fig \ref{fig_THMPWAdiargam1} describes the THM resonant amplitude  in the PWA.
For the process under consideration in Fig. \ref{fig_THMPWAdiargam1},  $\;\;a={}^{14}{\rm N},\,\;
A={}^{12}{\rm C},\,\;s=d, \,\; F={}^{24}{\rm Mg}^{*},$ $\;b$ and $B$ are the nuclei in the two-fragment channel into which the resonance ${}^{24}{\rm Mg}$ decays. 
This diagram describes the process in which the $a-A$ relative motion in the initial channel of the reaction and $b-F$ in the final state is described by the plane waves.

\begin{figure}
\includegraphics[width=3.0in,height=4.5in]{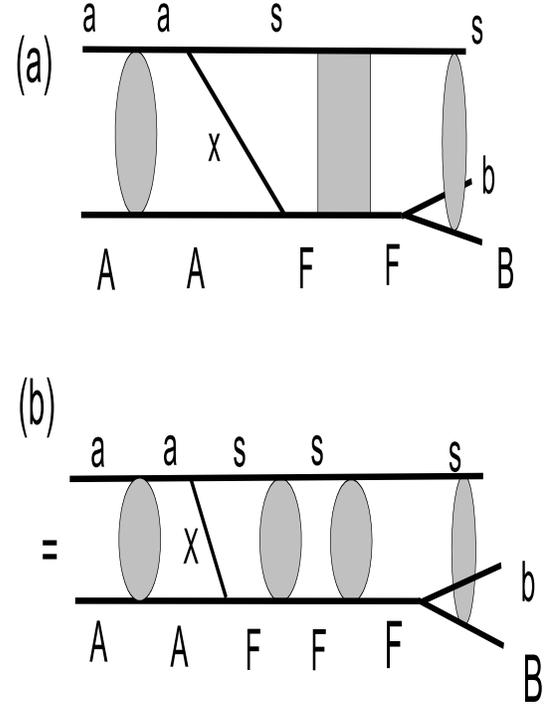}
  \caption{The diagrams describing the THM mechanism including the Coulomb interactions in the initial, final and intermediate states. In the diagrams, the grey bulb on the left side is the Coulomb $a-A$ scattering in the initial channel described by the Coulomb scattering wave function. The grey rectangle in the top diagram is the Green function in the final state describing the propagation of the system $s+F$, where $F$ is the resonance. The grey bulb on the right side describes the intermediate srate three-body Coulomb interaction given by the three-body Coulomb wave function. In the bottom diagram, the Green function is replaced by its spectral decomposition, which includes the Coulomb scattering wave functions (grey bulbs) describing the Coulomb rescattering 
$s$ in the intermediate state.}
\label{fig_THMCouldiagram1}
\end{figure}
A correct diagram, which replaces the simple PWA one in Fig. \ref{fig_THMPWAdiargam1}, is given in Fig. \ref{fig_THMCouldiagram1}.
The diagrams in Fig. \ref{fig_THMCouldiagram1} can be derived using a few-body approach and the detailed derivation will be given in the following up publication. It is important that in the intermediate state appears the Green function resolvent
\begin{align}
G(z)= \frac{1}{z - K - U_{sF}^{C} - V_{F} + i0},
\label{Greenfunct1}
\end{align}   
where  $K$ is the total kinetic energy operator of the system $s+F$ including the kinetic energy operators of the internal motion in $F$, $\;U_{sF}^{C}$ is the channel Coulomb potential describing the interaction between the c.m. of $s$ and $F$, $V_{F}$ is the internal interaction potential of $F$. The spectral decomposition of this Green function is expressed in terms of the $s+ F$ Coulomb scattering wave functions in the intermediate state and the complete set 
of the internal wave functions of $F$  from which the resonance term can be singled out by deformation of the integration contour. 
 
The THM mechanism is a two-step process. The first step is the transfer reaction $a+ A  \to s + F$ sandwiched by the Coulomb distorted waves in the initial and intermediate states, see the bottom diagram in Fig \ref{fig_THMCouldiagram1}. The second stage describes the decay of the resonance $F \to b+B$ sandwiched by the intermediate Coulomb distorted wave and the final-state three-body Coulomb wave function.

After very tedious transformations the transfer reaction amplitude can be singled out explicitly under the integral sign. This amplitude can be calculated using the standard   DWBA. The second stage describing the decay of the resonance sandwiched by the Coulomb scattering wave function and the final-state three-body Coulomb wave function can be calculated analytically using the technique developed in \cite{muk1985}. After some simplifications, the expression for the THM reaction amplitude can be reduced to

\begin{widetext}
\begin{align}
&M^{THM} \approx e^{-\pi\,\eta_{0}/2}\,\Gamma(1-i\,\eta_{0})\,\sqrt {\frac{{\pi\,{\Gamma _{bB}}}}{{{\mu _{bB}}k_{(R)bB}^{(0)}}}}\,e^{i\,\delta^{pot}(k_{(R)bB}^{0})}\,(2\,q_{0})^{-2i\,\eta_{0}}\,M^{DW}(q_{0}{\rm {\bf {\hat k}}}_{sF},\,{\rm {\bf k}}_{aA})
\nonumber\\
&\times \int \frac{{d{{\rm {\bf p}}_b}}}{{{{(2\pi )}^3}}}\frac{{d{{\rm {\bf p}}_B}}}{{{{(2\pi )}^3}}}\Phi _{{{{\bf{k}}_b},{\rm {\bf k}}_B}}^{( + )}({{\bf{p}}_b},{{\rm{\bf p}}_B})\,\frac{1}{{{{(p_{sF}^2 - q_{0}^2)}^{1 - i{\eta _{0}}}}}}.   
\label{Finalequation1}
\end{align} 
\end{widetext}

Here, $M^{DW}(q_{0}{\rm {\bf {\hat k}}}_{sF},\,{\rm {\bf k}}_{aA})$ is the amplitude of the  $a+ A \to s+ F$ reaction, which is the first stage of the THM reaction, and can be calculated using the DWBA,
$\delta^{pot}(k_{(R)bB}^{0})$ is the $b-B$ nonresonant scattering phase shift in the partial wave corresponding to the resonance under consideration (the index of the orbital angular momentum of the resonance, for simplicity, is omitted). In the c.m. of the THM reaction, $\,{\rm {\bf k}}_{s} +{\rm {\bf k}}_{b}+ {\rm {\bf k}}_{B}=0$,
$\,{\rm {\bf k}}_{i}$ and ${\rm {\bf p}}_{i}$ are the on-the -energy-shell (ONES) and off-the-energy-shell (OFES) momenta of particle $i$, 
\begin{align} 
q_{0}=(2\,\mu_{sF}[E_{int} - E_{(R)xA} ])^{1/2}.
\label{q01}
\end{align}
Here, $\,E_{int} = E_{sF} + E$, where $\,E \equiv E_{xA}$,  is the total kinetic energy in the intermediate state $\,s+ x+A\,$,  $\mu_{ij}$ is the reduced mass of particles $i$ and $j$, $\;{\rm {\bf k}}_{ij}$ and $\,{\rm {\bf p}}_{ij}$ are ONES and OFES relative momentum of particles $i$ and $j$, $\;\eta_{0}= (Z_{s}\,Z_{F}/137)\,(\mu_{sF}/q_{0})$, $\;\eta_{R}^{(0)}= (Z_{b}\,Z_{B}/137)\,(\mu_{bB}/k_{(R)bB}^{(0)})$, $\;Z_{i}e$ is the charge of particle $i$. $\Phi _{_{{{\bf{k}}_b},{\rm {\bf k}}_B}}^{( + )}({{\bf{p}}_b},{{\rm{\bf p}}_B})$ is the final-state three-body Coulomb scattering wave function in which the Coulomb $b-B$ interaction is switched off because it has already been taken into account in the decay vertex $F \to b+B$. At $E= E_{(R)xA}^{(0)}$  $\;q_{0}=(2\,\mu_{sF}[E_{sF} + i\,\Gamma/2)^{1/2}$.

The integral in  Eq. (\ref{Finalequation1}) depends on ${\rm {\bf k}}_{b}$ and ${\rm {\bf k}}_{B}$.
Hence the amplitude $M^{THM}$ depends on the three-body kinematics of the THM experiment. Because we do not 
know it, the three-body wave function $\Phi _{ {\rm{\bf k}}_b,{\rm {\bf k}}_B }^{( + )}({\rm{\bf p}}_b,{\rm{\bf p}}_B)$  is replaced by the two-body Coulomb scattering wave function  $\Psi_{ {\rm {\bf k}}_{sF} }^{C(+)}({\rm {\bf p}}_{sB}$ describing the final-state $s-F$ scattering wave function via the Coulomb channel potential $U_{sF}^{C}$ depending on the distance between the c.m. of nuclei $s$ and $F$. Note that $U_{sF}^{C}$ can be replaced by the sum of the Coulomb channel potential and the nuclear optical potential.
 The  two-body wave function $\Psi_{ {\rm {\bf k}}_{sF} }^{C(+)}({\rm {\bf p}}_{sB})$  allows us to simplify  the calculations significantly. However, by using  the two-body Coulomb scattering wave function rather than the three-body one we lose the final-state three-body-effects leading to acceleration (deceleration) of the  particles in the final  state.  
 
 Equation (\ref{Finalequation1}) can be  reduced 
  to
\begin{widetext}
\begin{align}
&M^{THM} \approx e^{-\pi\,\eta_{0}/2}\,\Gamma(1-i\,\eta_{0})\,\sqrt {\frac{{\pi\,{\Gamma _{bB}}}}{{{\mu _{bB}}k_{(R)bB}^{(0)}}}}\,e^{i\,\delta^{pot}(k_{(R)bB}^{0})}\,(2\,q_{0})^{-2i\,\eta_{0}}\,M^{DW}(q_{0}{\rm {\bf {\hat k}}}_{sF},\,{\rm {\bf k}}_{aA})
\nonumber\\
&\times \int \frac{{d{{\rm {\bf p}}_{sF}}}}{{{{(2\pi )}^3}}}\,\Psi _{ {\rm {\bf k}}_{sF} }^{C( + )}({\rm {\bf p}}_{sF})\,\frac{1}{{{{(p_{sF}^2 - q_{0}^2)}^{1 - i{\eta _{0}}}}}}.   
\label{Finalequation1}
\end{align} 
\end{widetext}

\begin{figure}[htbp]
\includegraphics[width=0.5\textwidth]{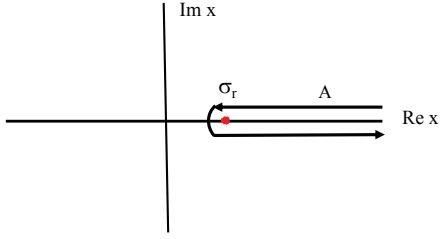}
\caption{Integration contour in the $x$-plane in Eq. (\ref{inttransf1}).  }
\label{fig_intcontour1}
\end{figure}

To calculate this integral  we use the Cauchy's theorem:
\begin{align}
\frac{1}{  \sigma_{r}^{ 1 - i\,\eta _{0}}  }= - \frac{1}{2\,\pi\,i}\,\oint\limits_{\sigma_{r}}^{\infty}{\rm d}x\,x^{-1+i\,\eta_{0}}\,\frac{1}{  (\sigma_{r}  -x )}
\label{inttransf1}
\end{align}
This integral is taken along a closed contour around $\sigma_{r}$ so that its integrand does not include any other singularities except for the pole at $x=\sigma_{r}$.
This integral can be deformed into a closed contour $A$, see Fig \ref{fig_intcontour1}, which begins at $x=\infty$, encircles the pole of the integrand at $x=\sigma_{r}$  and goes back to $x=\infty$.

Now we can use Eq. (1) from \cite{Nordsieck} :
\begin{align}
&\int\,\frac{   {\rm d}{\rm {\bf p}}}{{2\,\pi}^{3}}\,\Psi_{ {\rm {\bf k}}_{sF} }({\rm {\bf p}}_{sF}) \,\frac{1}{  p_{sF}^2 - q_{0}^2 -x }                           \nonumber\\
& = e^{-\pi\,\eta_{sF}/2}\, \Gamma(1+i\,\eta_{sF})\,
\frac{[-(k_{sF}+ i\,\sqrt{-q_{0}^{2} -x} )^{2}]^{i\,\eta_{sF}}}{[k_{sF}^{2}  - q_{0}^{2} -x ]^{1+i\,\eta_{sF}} }.
\label{Nordsieck1}
\end{align}
Then we get
\begin{widetext}
\begin{align}
J_{f}=\int \frac{{d{{\rm {\bf p}}_{sF}}}}{{{{(2\pi )}^3}}}\,\Psi _{ {\rm {\bf k}}_{sF} }^{C( + )}({\rm {\bf p}}_{sF})\,\frac{1}{{{{(p_{sF}^2 - q_{0}^2)}^{1 - i{\eta _{0}}}}}}=
 -e^{-\pi\,\eta_{sF}/2}\, \Gamma(1+i\,\eta_{sF})\, \frac{1}{2\,\pi\,i}\,\oint\limits_{\sigma_{r}}^{\infty}{\rm d}x\,x^{-1+i\,\eta_{0}}\,\frac{[-(k_{sF}+ i\,\sqrt{-q_{0}^{2} -x} \,)^{2}]^{i\,\eta_{sF}}}{[\sigma_{r} -x ]^{1+i\,\eta_{sF}} },
\label{int1}
\end{align}
\end{widetext}
where $\sigma_{r}= k_{sF}^{2}- q_{0}^{2}$.

Using the substitution  $x=\sigma_{r} \,y\;$ we can rewrite Eq. (\ref{int1})  as
\begin{align}
&J_{f}= -e^{-\pi\,\eta_{sF}/2}\, \Gamma(1+i\,\eta_{sF})\, \sigma_{r}^{i(\eta_{0} - \eta_{sF})-1}                \nonumber\\
&\times  \frac{1}{2\,\pi\,i}\,\oint\limits_{1}^{\infty}{\rm d}y\,y^{-1+i\,\eta_{0}}\,\frac{[-(k_{sF}+ i\,\sqrt{-q_{0}^{2} -\sigma_{r}\,y} )^{2}]^{i\,\eta_{sF}}}{[1 -y ]^{1+i\,\eta_{sF}} }.
\label{int2}
\end{align}
Because we consider $\sigma_{r} \to 0$ we can  take  the factor $[-(k_{sF}+ i\,\sqrt{-q_{0}^{2} -\sigma_{r}\,y} )^{2}]^{i\,\eta_{sF}}$ out under from the integral sign at $\sigma_{r}=0$.
Then Eq. (\ref{int2}) reduces to
\begin{widetext}
\begin{align}
J_{f}= -e^{-\pi\,\eta_{sF}/2}\, \Gamma(1+i\,\eta_{sF})\, \sigma_{r}^{i(\eta_{0} - \eta_{sF})-1} \, [-\Big( k_{sF}+ i\,\sqrt{ -q_{0}^{2} }\,\Big)^{2}]^{i\,\eta_{sF}}             
 \frac{1}{2\,\pi\,i}\,\oint\limits_{1}^{\infty}{\rm d}y\,y^{-1+i\,\eta_{0} }\,\frac{1}{    (1 -y )^{ 1+i\,\eta_{sF} } }.    
\label{int3}
\end{align}
\end{widetext}
The integrand in Eq. (\ref{int3})    has two cuts, from $y=1$ to $\infty$   and from $y=0$ to $-\infty$, see Fig. {\ref{fig_intcontour2}.  
\begin{figure}[htbp]
\includegraphics[width=0.5\textwidth]{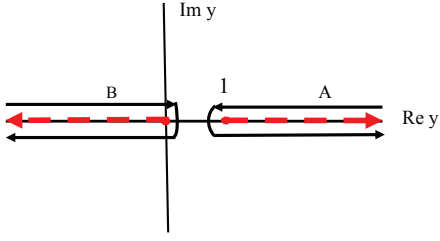}
\caption{Integration contours in the $y$-plane in Eq. (\ref{int3}).  The red dashed lines are the cuts from $y=0$ to $\infty$ and  from $y=0$ to $\infty$.. }
\label{fig_intcontour2}
\end{figure}
The  contour $A$  can deformed into the integration contour $B$, see Fig. \ref{fig_intcontour2}. 
The new integral  can be transformed  to
\begin{widetext}
\begin{align}
&J_{f} = -e^{-\pi\,\eta_{sF}/2}\, \Gamma(1+i\,\eta_{sF})\, \sigma_{r}^{i(\eta_{0} - \eta_{sF})-1} \, [-\Big( k_{sF}+ i\,\sqrt{ -q_{0}^{2} }\,\Big)^{2}]^{i\,\eta_{sF}}             
 \frac{1}{2\,\pi\,i}\,\oint\limits_{0}^{-\infty}{\rm d}y\,y^{-1+i\,\eta_{0} }\,\frac{1}{    (1 -y )^{ 1+i\,\eta_{sF} }       }     \nonumber\\
 &=i\,e^{-\pi\,\eta_{sF}/2}\, \Gamma(1+i\,\eta_{sF})\, \sigma_{r}^{i(\eta_{0} - \eta_{sF})-1} \, [-\Big( k_{sF}+ i\,\sqrt{ -q_{0}^{2} }\,\Big)^{2}]^{i\,\eta_{sF}}             
sh(\pi\,\eta_{0}) \,\int\limits_{0}^{\infty}{\rm d}y\,y^{-1+i\,\eta_{0} }\,\frac{1}{    (1 +y )^{ 1+i\,\eta_{sF} }   }  .
 \label{int4}
 \end{align}
 \end{widetext}
 
Taking into account Eq. (3.194(3))  from \cite{GR1} we get
\begin{align}
& J_{f}=i e^{-\pi\,\eta_{sF}/2}\, \Gamma(1+i\,\eta_{sF})\, \sigma_{r}^{i(\eta_{0} - \eta_{sF})-1}                \nonumber\\
& \times [-\Big( k_{sF}+ i\,\sqrt{ -q_{0}^{2} }\,\Big)^{2}]^{i\,\eta_{sF}}\,                                                   \nonumber\\
& \times sh(\pi\,\eta_{sF})\,\frac{\Gamma(i\,\eta_{0})\,\Gamma(1+ i\,[\eta_{sF}- \eta_{0}])}{\Gamma(1+i\,\eta_{sF})}.
\label{int5}
\end{align}
Recalling that  (see Eq. (8.332(3) from \cite{GR1})
\begin{align}
\Gamma(1- i\,\eta_{0})\,\Gamma(i\,\eta_{0})= -\frac{i\,\pi}{sh(\pi\,\eta_{0})},
\label{GG1}
\end{align}
and using  Eq. (\ref{int5}) we can simplify  Eq. (\ref{Finalequation1}) to
\begin{widetext}
\begin{align}
&M^{THM} \approx \frac{ N(E, E_{(R)xA} )}{ (k_{sF}^2 - q_{0}^{2}) } \,\pi\,\sqrt {\frac{{\pi\,{\Gamma _{bB}}}}{{{\mu _{bB}}k_{(R)bB}^{(0)}}}}\,e^{i\,\delta^{pot}(k_{(R)bB}^{0})}\,M^{DW}(q_{0}{\rm {\bf {\hat k}}}_{sF},\,{\rm {\bf k}}_{aA}) ,   
\label{Finalequation2}
\end{align} 
\end{widetext}
where 
\begin{align}
&N(E_{xA}, E_{(R)xA} )= e^{\pi\,(\eta_{sF}-\eta_{0})/2}\,\Gamma(1+i\,[\eta_{sF}-\eta_{0}])\,                                                               \nonumber\\
& \times \frac{   [\Big( k_{sF}+ i\,\sqrt{ -q_{0}^{2} }\,\Big)^{2}]^{i\,\eta_{sF}}  }{    (k_{sF}^2 - q_{0}^{2})^{ i(\eta_{SF}-\eta _{0}) }  }\, \frac{1}{(2\,q_{0})^{2i\,\eta_{0}}}. 
\label{N1}
\end{align}

Taking into account that 
\begin{align}
& \lim_{E - E_{(R)xA}^{(0)} \to 0,\,\Gamma/(8\,E_{sF}) \to 0} \frac{[\Big( k_{sF}+ i\,\sqrt{ -q_{0}^{2} }\,\Big)^{2}]^{i\,\eta_{sF}}}{(2\,q_{0})^{2i\,\eta_{0}}}   =\frac{1}{ (2\,k_{sF} )^{i\,\eta_{sF}}}
\label{factor1}
\end{align}
we get  that at $E \to E_{(R)xA}^{(0)}$
\begin{align}
&N(E, E_{(R)xA} )= \frac{ 1 }{[ 2\,\mu_{sF}(E_{(R)xA}^{(0)} - E +i\,\frac{\Gamma} {2})]^{ i(\eta_{sF}-\eta _{0}) }  }                     \nonumber\\
& \times \frac{1}{ (2\,k_{sF} )^{i\,\eta_{sF}}} . 
\label{N1}
\end{align}
and 
\begin{align}
 \lim_{E - E_{(R)xA}^{(0)} \to 0,\,\Gamma/(8\,E_{sF}) \to 0} |N(E,E_{(R)xA)}|  =1,
 \label{modulN1}
 \end{align}
because $\lim (\eta_{0} -\eta_{sF}) =0 $ at these conditions.

We can write the final expression for theTHM amplitude:
\begin{widetext}
\begin{align}
M^{THM}\approx \frac{ {\sqrt{\Gamma _{bB} }}}{ E_{(R)xA}^{(0)} -E - i\frac{\Gamma }{2} }\,\sqrt {\frac{{\pi}}{{{\mu _{bB}}k_{(R)bB}^{(0)}}}}\,e^{i\,\delta^{pot}(k_{(R)bB}^{0})}\,\frac{\pi\, N(E, E_{(R)xA} )}  { (2\,k_{sF} )^{i\,\eta_{sF}}}\,M^{DW}({\rm {\bf  k}}_{sF}, {\rm {\bf k}}_{aA}). 
\label{Finaleq3}
\end{align}
\end{widetext}

One of the important features of Eq. (\ref{Finaleq3}) is that the resonance pole becomes a branching point singularity because of the Coulomb interaction of the spectator $s$ with the resonance $F$ in the intermediate state.
A  priori, the branching point singularity can change the shape, location and the strength of the resonance. It can be seen for wider resonances. Here we assume that the resonance is narrow.

Equation (\ref{Finaleq3}) is not yet the conventional THM amplitude. To get the THM amplitude one needs  to single out from Eq. (\ref{Finaleq3}) the resonant S-matrix $\,P_{l_{xA}}^{-1}{\cal S}_{xA \to bB}$ of the binary resonant sub-reaction $x+ A \to b+B$  in which the penetrability factor $P_{l_{xA}}$ in the entry channel  $x+ A$ is excluded. 
The reason for the absence of the penetrability factor $P_{l_{xA}}$  in the entry channel of the binary sub-reaction of the THM reaction is explained in \cite{muk2011}, see Eq. (117) 
of  \cite{muk2011}. 

Now using the surface integral formalism the $M^{DW}$ amplitude can be transformed into the surface term and peripheral one while the internal part becomes small (see Eq. (117) \cite{muk2011}). The surface and external terms are parametrized in terms of the reduced width $\gamma_{xA\,l_{xA}}$ of the entry channel x+A of the resonant binary subreaction 
$x+ A  \to F \to b +B$. Then Eq. (\ref{Finaleq3}) can be rewritten in the THM form:
\begin{align}
&M^{THM} \approx \frac{1}{\sqrt{2\,P_{l_{xA}}}}\,\frac{{\sqrt {\Gamma_{xA}{\Gamma _{bB}}} }}{E - E_{xA} - i\frac{\Gamma }{2}}\,\sqrt {\frac{{\pi}}{{{\mu _{bB}}k_{(R)bB}^{(0)}}}}\,                 \nonumber\\
& \times e^{i\,\delta^{pot}(k_{(R)bB}^{0})}\,\frac{\pi\, N(E, E_{(R)xA} )}  { (2\,k_{sF} )^{i\,\eta_{sF}}}\,{\cal M}^{DW}({\rm {\bf  k}}_{sF}, {\rm {\bf k}}_{aA}). 
\label{THMfinal}
\end{align}
Here ${\cal M}^{DW}$ is the DWBA amplitude in the surface integral representation from which the reduced width amplitude is singled out \cite{muk2011}. To simplify the notations we dropped all the spin-angular momentum dependences, which are explicitly written down in Eq. (117) of \cite{muk2011}.
The  astrophysical factor for the resonant reaction is  given by Eq. (24) of \cite{muk2017}.

The THM triple DCS is expressed in terms of the THM reaction amplitude $M^\textrm{THM}$ (we need to take into account the spin-angular momentum dependence of this amplitude, which can be recovered using Eq. (117) of Ref.\cite{muk2011}) by Eq. (33) of Ref. \cite{muk2017}. After integration over the solid angle $\Omega_{{\rm {\bf k}}_{bB}}$ we get an important relationship between the THM double DCS and the astrophysical factor for the THM resonant reaction proceeding through an isolated resonance: 
\begin{align}
&\frac{{{{\rm{d}}^2}{\sigma^\textrm{THM}}}}{{{\rm{d}}E{\mkern 1mu} {\rm{d}}{\Omega _{{{\bf{k}}_{sF}}}}}} = K(E)\,S(E){\mkern 1mu}\,\frac{{{\rm{d}}{\sigma^\textrm{DW}}(E,\cos {\theta _s})}}{{{\rm{d}}{\Omega _{{{\bf{k}}_{sF}}}}}}{|N(E){|^2}}.
\label{doubleTHMcrSfactor1}
\end{align}
Here,  $\frac{{{\rm{d}}{\sigma^\textrm{DW}}(E,\cos {\theta _s})}}{{{\rm{d}}{\Omega _{{{\bf{k}}_{sF}}}}}}$ is the DWBA DCS of the  ${}^{14}{\rm N} + {}^{12}{\rm C} \to d+ {}^{24}{\rm Mg}^{*}$ reaction populating the isolated resonance state, $\theta_{s}$ is the scattering angle of the spectator $s$ (deuteron) in the c.m. of the THM reaction, $S(E)$ is the astrophysical factor.  
$\,K(E)$ is a trivial kinematical factor whose explicit expression is not important for our purposes.
Because we have shown that  for narrow resonances and $E=E_{(R)xA}^{(0)}$  $\;|N(E,E_{(R)xA})|= 1$, from now on we omit the factor $|N(E,E_{(R)xA})|^{2}$ . It should be underscored that this factor can be omitted from the THM DCS only because we replaced the final-state three-body Coulomb wave function by the two-body one.  The inclusion of the three-body wave function leads to a more complicated expression for the function $N(E,E_{(R)xA})$. To calculate this function we should know  the complete three-body kinematics of  an experiment.

The THM astrophysical factor  determined from  Eq. (\ref{doubleTHMcrSfactor1}) is
\begin{align}
&S(E) = N_{F}\,K(E)\,\frac{{{{\rm d}^2}{\sigma^\textrm{THM}}}}{{{\rm d}E\,{\rm d}{\Omega _{{{\rm {\bf k}}_{sF}}}}}}\frac{1}{{\frac{{{\rm d}{\sigma^\textrm{DW}}(E,\cos {\theta _s})}}{{{\rm d}{\Omega _{{{\rm {\bf k}}_{sF}}}}}}}}.
\label{SfactorarbE1}
\end{align}
Here, $N_{F}$ is an overall, energy-independent  normalization factor of the THM data to direct data. We remind to the reader that to get the absolute value of the THM $S(E)$-factor one needs to normalize it to the direct data available at higher energies.

Equations (\ref{doubleTHMcrSfactor1}) and (\ref{SfactorarbE1}) are pivotal for understanding of the problem of extraction of the $S(E)$-factor from the THM DCS. Because in the normalization interval of $E=2.5-2.66$ MeV the outgoing deuterons are below the Coulomb barrier,  $\frac{{{\rm{d}}{\sigma^\textrm{DW}}(E,\cos {\theta _s})}}{{{\rm{d}}{\Omega _{{{\bf{k}}_{sF}}}}}}$  is small and rapidly increases when the resonance energy $E$ decreases. This increase of $\frac{{{\rm{d}}{\sigma^\textrm{DW}}(E,\cos {\theta _s})}}{{{\rm{d}}{\Omega _{{{\bf{k}}_{sF}}}}}}$ should be reflected in the behavior of $\frac{{{{\rm{d}}^2}{\sigma^\textrm{THM}}}}{{{\rm{d}}E{\mkern 1mu} {\rm{d}}{\Omega _{{{\bf{k}}_{sF}}}}}}$ and the THM $S(E)$-factor. As we mentioned, in Ref. \cite{Nature} instead of the DWBA a simple PWA was used. 
In the PWA the distorted waves in the initial and final states of the transfer reaction are replaced by the corresponding plane waves.
The DCS as a function of $E_{sF}$ obtained using the PWA changes very little compared to the change of the DWBA DCS. This is the main reason why the THM astrophysical factors in \cite{Nature} show unusually high rise when $E$ decreases. 
 
The normalization interval selected in \cite{Nature} was chosen to be $E=2.5-2.63$ MeV. However, there are two resonances with negative parities, which are questionable because two colliding  identical bosons ${}^{12}{\rm C} + {}^{12}{\rm C}$ cannot populate resonances with negative parity. There are two resonances with positive parities cited in \cite{Nature}: at $2.664$ and $2.537$ MeV. It was underscored  in \cite{Nature} that the THM data reproduce the higher-lying resonance. That is why it is assumed here that the normalization factor $N_{F}$ is determined by the normalization of the THM astrophysical factor to the directly measured resonance at $E=2.664$ MeV. Practically we selected the normalization of the THM data on the edge of the energy interval measured in \cite{Nature}. 

We remind now that in the PWA the THM astrophysical factor for an isolated resonance is given by (see Eq. (35) from Ref. \cite{muk2017}
\begin{align}
&S^\textrm{(PWA)}(E) = N_{F}\,K(E)\,\frac{{{{\rm d}^2}{\sigma^\textrm{THM}}}}{{{\rm d}E\,{\rm d}{\Omega _{{{\rm {\bf k}}_{sF}}}}}}\frac{1}{\varphi_{a}^{2}\,|M_{l_{xA}}|^{2}}.
\label{SPWA1}
\end{align}
The factor $M_{l_{xA}}$ was obtained in \cite{muk2011}, $\varphi_{a}$ is the $a=(s\,x)$ (here, ${}^{14}{\rm N}=(d\,{}^{12}{\rm C})$) bound-state wave function.

To compare Eqs. (\ref{SfactorarbE1})  and  (\ref{SPWA1}) the factor $M_{l_{xA}}$ is singled out from the DWBA DCS. As a result we obtain that
\begin{align}
\frac{{{\rm{d}}{\sigma^\textrm{DW}}(E,\cos {\theta _s})}}{{{\rm{d}}{\Omega _{{{\bf{k}}_{sF}}}}}}=
|M_{l_{xA}}|^{2}\,\frac{{{\rm{d}}{\sigma^\textrm{DWZR}}(E,\cos {\theta _s})}}{{{\rm{d}}{\Omega _{{{\bf{k}}_{sF}}}}}},
\label{DWBAZR1}
\end{align}
where $\frac{{{\rm{d}}{\sigma^\textrm{DWZR}}(E,\cos {\theta _s})}}{{{\rm{d}}{\Omega _{{{\bf{k}}_{sF}}}}}}$ is the zero-range (in the vertex $x+ A \to F$) DWBA DCS for the resonant reaction $a+ A \to s+F$ (here, ${}^{14}{\rm N} + {}^{12}{\rm C} \to .d + {}^{24}{\rm Mg}^{*}$). 

The renormalization factor of the THM astrophysical factor is obtained by  taking the ratio of the $S(E)$ factors given by Eqs (\ref{SfactorarbE1}) and (\ref{SPWA1}) :
\begin{align}
&R(E)=  \frac{{{\rm{d}}{\sigma^\textrm{DWZR}}(2.664\,\,{\rm{MeV}},\cos {\theta _s})/{\rm{d}}{\Omega _{{{\bf{k}}_{sF}}}}}}{{{\rm{d}}{\sigma^\textrm{DWZR}}(E,\cos {\theta _s})/{\rm{d}}{\Omega _{{{\bf{k}}_{sF}}}}}}.
\label{RF1}
\end{align}
The renormalization factor appears due to the inclusion of the Coulomb distorted waves in the initial and finals states of the ${}^{12}{\rm C}$ transfer reaction  of the THM reaction. The DWBAZR DCSs are calculated using the FRESCO code \cite{FRESCO}. Note that for comparison we also calculated ${{\rm{d}}{\sigma^\textrm{DWZR}}(E,\cos {\theta _s})/{\rm{d}}{\Omega _{{{\bf{k}}_{sF}}}}}$ including the Coulomb plus nuclear distorted waves. The bound-state wave function $\varphi_{a}^{2}$ can be dropped because it does not depend on energy. The renormalization factor $R(E)$ is set equal to unity at the THM normalization energy of $E=2.664$ MeV.

Results of the calculations are presented in  Fig. \ref{fig_RS(p)}. The  renormalized astrophysical factors are $R(E)\,S^{*}(E)$, where $\,S^{*}(E)$ are taken from \cite{Nature} (here we use the notation $S^{*}(E)$ for the astrophysical factor, which was used in \cite{Nature}). In panel (a) is shown the behavior of the astrophysical factors for the channel $p_{0} + {}^{23}{\rm Na}$. Because a similar behavior of the $R(E)\,S^{*}(E)$-factors is found for three other channels, $p_{1}+ {}^{23}{\rm Na}(0.44\, {\rm MeV})$, $\;\alpha_{0} + {}^{20}{\rm Ne}$ and $\;\alpha_{1} + {}^{20}{\rm Ne}(1.63\, {\rm MeV})$, in panel (b) we show the total astrophysical factors, which are given by the sum of the astrophysical factors of four final channels which were detected in \cite{Nature}. We find that at the resonance energies $E=0.8-0.9$ MeV the renormalization factor $R(E)$ decreases the THM astrophysical factors from \cite{Nature}  by a factor of $\approx 10^{3}$. 

\begin{figure}[htbp]
\includegraphics[width=0.5\textwidth]{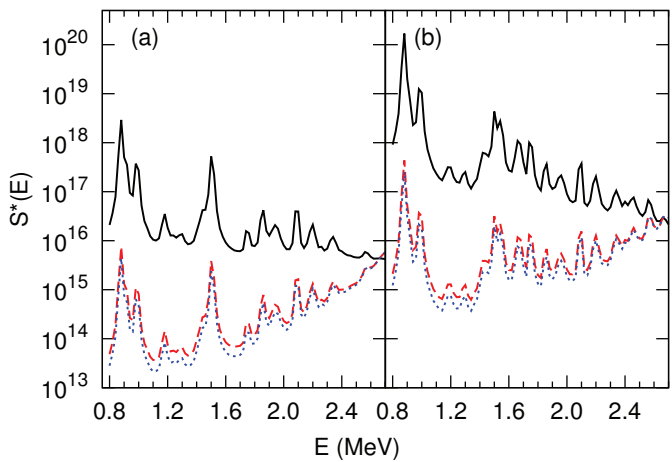}
\caption{$S^{*}(E)$-factors for the ${}^{12}{\rm C}+ {}^{12}{\rm C}$ fusion. Panel\,(a): $S^{*}(E)$-factors for the reaction ${}^{12}{\rm C} + {}^{12}{\rm C} \to {p_{0}} + {}^{23}{\rm Na}$. Black solid line is the $S^{*}(E)$-factor from \cite{Nature}; the blue dotted line is the renormalized $R(E)\,S^{*}(E)$-factor calculated using the pure Coulomb distorted waves; the magenta dash line is the renormalized $R(E)\,S^{*}(E)$-factor calculated using the Coulomb plus nuclear distorted waves. Panel\,(b): The total astrophysical factors for the ${}^{12}{\rm C}+ {}^{12}{\rm C}$ fusion. The notations are the same as in panel (a).}
\label{fig_RS(p)}
\end{figure}

In Ref. \cite{arxivAurora},  the authors extrapolated our renormalized astrophysical factors to the energies higher then the normalization energy at $E=2.664$ MeV to prove that the inclusion of the distorted waves is not valid. First, the extrapolations shown in \cite{arxivAurora}  are not legitimate because to find the extrapolation factor at $E>2.664$ MeV one must have the THM $S^{*}(E)$-factor, which is not available. Besides, the extrapolation in \cite{arxivAurora} is done to the region in which the THM reaction is forbidden because of the energy conservation. Nevertheless, if we use the normalization point at the energy $E< 2.664$ MeV, then the extrapolation of our renormalized $S(E)$-factor goes higher than the direct data at energies higher than the normalization energy. Because we just replaced the PWA by the DWBA (no doubt that the DWBA reflects physics better than the simple PWA) the extrapolation of our renormalized $R(E) \,S^{*}(E)$-factors  only confirms once again that there is an issue with the THM data, especially at higher resonance energies, and the astrophysical factors shown in \cite{Nature} are the results of the application of the PWA theory, which is invalid in the case under consideration.  Note that after the integration over  the $E$ of both sides of Eq. (\ref{doubleTHMcrSfactor1}) in the region which includes a selected narrow isolated resonance (let it ibe $E=E_{0}$) we obtain that the THM double DCS reduces to the single THM DCS. This THM DCS is proportional to the DCS of the carbon transfer reaction populating  the resonance  $E_{0}$. Our estimations show that this DCS is extremely small in the THM normalization interval and is very difficult to measure. 

To corroborate further our findings in Fig \ref{fig_renormS} we present the renormalization factors $R(E)$ at three different incident energies of ${}^{14}{\rm N}$: $30,\,33$ and $35$ MeV. The first energy is used in the THM experiment \cite{Nature}. We see a strong drop of $R(E)$ for $E_{{}^{14}{\rm N}}=30$ MeV . We do not discuss here whether the higher energies would allow one to cover the whole resonance energy interval. We just demonstrate that the renormalization  factor $R(E)$ quickly approaches unity when the incident energy of  ${}^{14}{\rm N}$ increases  confirming that the Coulomb interactions are the main reason for the drop of $R(E)$ at $30$ MeV. This again confirms that at this energy the simple PWA is not valid.
From Fig. \ref{fig_renormS}  we can see also that the main drop   of the renormalization factor, about a factor of $\approx 0.01$, occurs   when the energy  decreases from $2.66$ MeV to $2$ MeV, while it drops only by a factor of $ \approx 0.1$ when the energy decreases from $2$ MeV to $0.8$ MeV.

\begin{figure}[htbp]
\includegraphics[width=0.4\textwidth]{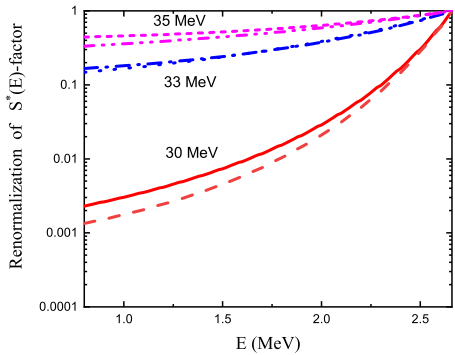}
\caption{Renormalization factors $R(E)$ calculated at three different incident energies of ${}^{14}{\rm N}$. Red lines are $R(E)$ for $E_{{}^{14}{\rm N}}=30$ MeV: solid line is calculated with pure Coulomb distorted waves, dashed line corresponds to the Coulomb plus nuclear distorted waves; blue lines are $R(E)$ for $E_{{}^{14}{\rm N}}=33$ MeV: dotted line is for the Coulomb distorted waves, dash-dotted line is for the Coulomb plus nuclear distorted waves; magenta lines are $R(E)$ for $E_{{}^{14}{\rm N}}=35$ MeV: dash-dotted-dotted line is for the Coulomb distorted wave, short dash line is for the Coulomb plus nuclear distorted waves.} 
\label{fig_renormS}
\end{figure}

The THM is a powerful and unique indirect method, which allows one to measure the $S(E)$-factors of the  resonant reactions down to astrophysically relevant energies, where direct methods are not able to obtain data due to very small cross sections.  Because the THM deals with three-body reactions rather than binary ones, a reliable theoretical analysis of the  THM data becomes critically important. For the THM reactions with the neutron-spectator or for the reactions with the energies above the Coulomb barrier and for interacting nuclei with small charges, the simple PWA works quite well and the THM results are expected to be reliable. However, this is not the case for the THM reaction under consideration, which aims to determine the astrophysical factors of the ${}^{12}{\rm C}+ {}^{12}{\rm C}$ fusion. In this reaction we deal with the strong Coulomb interactions in the  THM reaction. Moreover, the energies of the deuteron-spectator in the final state are below the Coulomb barrier. We have demonstrated here that the replacement of the PWA by the approach, which takes into account the Coulomb distortions, decreases the THM astrophysical factors from \cite{Nature} at the resonance energies of $E=0.8-0.9$ MeV by as much as $\approx 10^{3}$ times. Inclusion of the Coulomb plus nuclear distortions does not change this conclusion. We believe that the problem with the  astrophysical factors for the carbon-carbon fusion calls for new indirect experiments.

\section{Acknowledgments}
A. M. M. acknowledges the support by the U.S. DOE Grant No. DE-FG02-93ER40773, NNSA Grant No. DE-NA0003841
and U.S. NSF Award No. PHY-1415656. D.Y.P. acknowledges the support by the
national key research and development program (2016YFA0400502) and the 
support by NSFC Grant Nos. 11775013 and U1432247.
The authors thanks A. S. Kadyrov and X. D. Tang for help.

\end{document}